\documentclass[11pt]{article}
\usepackage{amsmath,amssymb,amsfonts,epsfig,graphicx,euscript}
\newcommand{\bse}{\begin{subequations}}
\newcommand{\ese}{\end{subequations}}
\newcommand{\be}{\begin{equation}}
\newcommand{\ee}{\end{equation}}
\newcommand{\bea}{\begin{eqnarray}}
\newcommand{\eea}{\end{eqnarray}}
\newcommand{\ba}{\begin{array}}
\newcommand{\ea}{\end{array}}

\begin{document}
\hfill%
\vbox{
    \halign{#\hfil        \cr
           \cr
                     }
      }
\vspace{1cm}
\begin{center}
{ \Large{\textbf{Strange metals at finite 't Hooft coupling }
\\}}
\vspace*{2cm}
{\bf Kazem Bitaghsir Fadafan$^1$}\\%
\vspace*{0.4cm}
{\it {$^2$Physics Department, Shahrood University of Technology,\\
P.O.Box 3619995161, Shahrood, Iran }}\\
{E-mails: {\tt bitaghsir@shahroodut.ac.ir}}%
\vspace*{1.5cm}
\end{center}

\vspace{.5cm}
\bigskip
\begin{center}
\textbf{Abstract}
\end{center}
In this paper, we consider the AdS-Schwarzshild black hole in
light-cone coordinates which exhibits non-relativistic z=2
Schrodinger symmetry. Then, we use the $AdS/CFT$ correspondence to
investigate the effect of finite-coupling corrections to two
important properties of the strange metals which are the Ohmic
resistivity and the inverse Hall angle. It is shown that the Ohmic
resistivity and inverse Hall angle are linear and quadratic
temperature dependent in the case of $\mathcal{R}^4$ corrections,
respectively. While in the case of Gauss-Bonnet gravity, we find
that the inverse Hall angle is quadratic temperature dependent and
the Ohmic conductivity can never be linear temperature dependent.
\newpage
\tableofcontents
\section{Introduction}

The AdS/CFT correspondence has been a powerful tool for studying
dynamics of strongly coupled field theory. Recently the application
of this duality in condensed matter physics (called $AdS/CMT$) has
been studied \cite{review1}. One of the most interesting subjects in
this content is to understand the strange metal behavior of heavy
fermion compounds and high temperature superconductors
\cite{strange1,strange2}. They show two important properties: linear
temperature dependent resistivity and quadratic temperature
dependent inverse Hall angle which can be expressed as%

\be \rho=(\sigma^{yy})^{-1}\,\sim\,T,\,\,\,\,\,\,\left(\cot
\theta_H\right)^{-1}\,\sim\,\frac{\sigma^{yy}}{\sigma^{yz}}\sim\,T^2.\label{strange}\ee
Because of the Hall current due to the magnetic field, $\theta_H$ is
called the Hall angle. The other property of strange metals is
related to the scaling of AC conductivity which we will not pursuit
in this paper. It is widely believed that to gain a better
understanding of such properties requires us to go beyond the regime
of weak coupling \cite{Sachdev:2010uj}.

To study non-relativistic phenomena in condensed matter, one should
investigate the non-relativistic generalization of the $AdS/CFT$
correspondence which has become an active research area
\cite{Son:2008ye,Balasubramanian:2008dm,Maldacena}. By introducing
the $Schr\ddot{o}dinger$ symmetry the $Schr\ddot{o}dinger$ space
successfully fitted into the AdS/CFT correspondence. A direct method
to obtain this space is to use the Null Melvin Twist to the known
solutions of type II supergravity \cite{Alishahiha:2003ru}. Another
way is to use the light-cone coordinates, then $Schr\ddot{o}dinger$
space can be found in the pure $AdS$ \cite{Goldberger:2008vg}.
Properties of Schr¨odinger black holes from those of AdS black holes
expressed in light-cone coordinates were discussed in
\cite{Kim:2010tf}.

Using the AdS/CFT, The strange metal behaviors were studied in
\cite{Faulkner:2010zz,Charmousis:2010zz,Myers:2010pk,Pal:2010sx,Lee:2010ii,Meyer:2011xn,Lee:2010uy}
. One should notice that there exists a few holographic systems
which show the particular temperature dependence of the Ohmic
resistivity and Hall angle in \eqref{strange}. One also finds that
the gravitational solutions showing the Lifshitz-like scaling does
not reproduce the behavior of the Hall angle \cite{pol}. A
holographic model building approach to 'strange metallic'
phenomenology was proposed in \cite{Pal:2010sx} and it was argued
that the spatial part of the metric components of background metric
should not be same.

In our study, we will consider $AdS$ space in the light-cone frame
(ALCF). This frame is proposed as a physical system which describes
the physics of strange metals \cite{KKP}. Comparing ALCF background
and a Lifshitz background was discussed in \cite{Kim:2011zd}. It was
shown that how an extra parameter can change the temperature scaling
behavior of conductivity.

Our purpose is to find the temperature dependence of \eqref{strange}
in the presence of higher derivative corrections. These corrections
on the gravity side correspond to finite-coupling corrections on the
gauge theory side. The main motivation to consider corrections comes
from the fact that string theory contains higher derivative
corrections arising from stringy effects. In the case of
$\mathcal{N}=4$ SYM theory, the leading order correction in
$1/\lambda$ arises from stringy correction to the low energy
effective action of type $\amalg$b supergravity, $\alpha'^3
{\cal{R}}^4$. The DC conductivity of massive $\mathcal{N} = 2$
hypermultiplet fields in an $\mathcal{N} = 4\, SU(N_c)$
super-Yang-Mills theory plasma in the large $N_c$ and finite 't
Hooft coupling was studied in \cite{AliAkbari:2010av}. Here, we
continue this study to investigate ${\cal{R}}^4$ and Gauss-Bonnet
corrections to two important properties of strange metals in
\eqref{strange}. An understanding of how the strange metal behaviors
affected by finite $\lambda$ corrections may be essential for
theoretical predictions.

The article is organized as follows. In the next section, we will
present the basic idea for studying the temperature dependence of
the resistivity in the ALCF framework. We follow the same direction
and consider the corrections to the strange metal behaviors in section 3.
In the last section we draw our conclusions and summarize our results. %

\section{Non-relativistic DC conductivity from AdS Light-Cone black hole}

In this section, we explore calculating the non-relativistic DC
conductivity in ALCF. The non-relativistic DC conductivity from the
Schr¨odinger black hole spacetime has been obtained in
 \cite{Ammon:2010eq}. They use the dual
gravitational description of probe D7 branes in an asymptotically
Schrodinger spacetime.

We consider the full 10-dimensional spacetime as $AdS_5$ times $S_5$
metric %
\be\label{metricform}%
 ds^2=G_{tt}\ dt^2+G_{xx}dx^2+G_{yy}dy^2+G_{zz}dz^2+G_{uu}\ du^2+
 d^2\Omega_5\,. %
\ee %
Here $(x,y,z,t)$ are field theory space and the metric functions are
given by $G_{tt}, G_{xx},\,G_{yy}, G_{zz}$ and $G_{uu}$. The radial
coordinates denotes by $u$. The $AdS$ boundary is located at
infinity and $u_0$ is event horizon. We will drop the $S_5$ part of
the metric for the rest of our discussion and discuss it when
introduce D7 branes on the background
\eqref{metricform}.\footnote{One should notice that in the
Schr¨odinger spacetime this part of the metric is very important
becasue of the non-trivial Kalb-Ramond field \cite{Ammon:2010eq}.}

The metric of $ALCF$ is obtained from \eqref{metricform} by the
transformation $x^{+}=b(t+x)$ and $x^{-}=\frac{1}{2b}(t-x)$, which
yields \cite{Kim:2010tf,Ammon:2010eq}%
\begin{align}
  ds^2 &= g_{++} dx^{+2} + 2 g_{+-} dx^+ dx^- + g_{--} dx^{-2}
  + \nonumber \\
  &G_{yy}dy^2+G_{zz}dz^2+G_{uu}\ du^2\,,
  \label{ALC}
\end{align}
where%
\begin{align}
 g_{++}=\frac{G_{tt}+G_{xx}}{4b^2},\,\,\,\,\,\,\,g_{+-}=\frac{G_{tt}-G_{xx}}{2},\,\,\,
  g_{--}=b^2(G_{tt}+G_{xx}).
\end{align}%
This coordinate transformation is introduced in \cite{Maldacena}.

The light-cone direction $x^+$ is identified as time and the
momentum of the $x^-$ direction is fixed. The symmetry is broken to
the Lifshitz $(z=2)$ symmetry. A key ingredient of ALCF is
transition from AdS $(z=1)$ to Lifshitz $(z=2)$. This transition is
controlled by the parameter $b$. In brief, the corresponding dual
theory interpolates between the usual relativistic scale symmetry
and the Lifshitz symmetry. The role and interpretation of the
parameter $b$ is explored in \cite{KKP}. One finds that
thermodynamic properties of the ALCF are identical to those of
Schr\"odinger space \cite{Kim:2010tf}.

One should introduce $N_f$ D7 branes on the background metric of
ALCF \eqref{ALC} to calculate the non-relativistic DC conductivity.
Notice that we work in the probe limit so that $N_f<<N_c$. Then we
ignore the quantum effects due to flavor fields. These D7-branes
fill $AdS_5$ and wrap the $S^3\subset S^5$. There are two remaining
world volume scalars on the branes. One scalar is chosen to be
trivially constant and the other a function of radial coordinate
$\theta(u)$ which describes the position $S^3$ on the $S^5$ and is
dual to the mass operator \cite{Karch}.

The $U(1)$ worldvolume gauge field $A_{\mu}$ is dual to the $U(1)$
current $J^\mu$. We follow the conventions of \cite{Kim:2010tf} and
turn on $A_+(y,u)$, $A_-(y,u)$ and $A_y(x^-,u)$ as well as the
$\theta(u)$. The above system is described by Dirac-Born-Infeld
action as follows%
\be S_{D7}=-N_{f} T_{D7}\int e^{-\phi}d^8 \xi
\sqrt{-det(g_{D7}+\tilde{F})}, \ee %
where $T_{D7}, \xi, \tilde{F}=(2\pi\alpha')F$ and $\phi$ are the
D-brane tension, worldvolume coordinates, the normalized $U(1)$
field strength and the dilaton, respectively. The pullback of the
spacetime metric with respect to the aforementioned embedding map is
given by the metric $g_{D7}$. We define $\mathcal{N}=2 \pi ^2 N_f
T_{D7}$ with $2 \pi^2$ from the integration of $S^3$ \footnote{In
$AdS_5\times S^5$ AdS black brane background, $e^{-\phi}$ is a
constant and it can be absorbed in ${\cal{N}}$.}.

From the gauge field equations of motion, conserved
charges $I_+=<J^+>$, $I_-=<J^->$ and $I_y=<J^y>$ associated with
$A_+(y,u)$, $A_-(y,u)$ and $A_y(x^-,u)$ are found. Following \cite{Kim:2010tf}, we introduce the gauge filed $A_\mu $ in the light-cone coordinates as%
\begin{align}
  A_+=E_b y + h_+(u),\,\,\,\,\, A_-=2 b^2 E_b y+h_-(u),\,\,\, A_y=A_y(u),\,\,\,A_u=0.\label{AAA}
\end{align}%
where the electric field is redefined as $E_b=\frac{E}{2b}$ to scale
non-relativistically and it is the light-cone component of an
electric field in the boost direction x \cite{Ammon:2010eq}. Having
worked out the gauge
fields in terms of $I_+,\,I_-$ and $E_b$, one finds the on-shell DBI action as follows%
\begin{align}%
S_{D7}=-\tilde{\mathcal{N}}^2\, \int du\,
\hat{G}_{1}(u)\,\sqrt{G_{+-}\,g_{uu}^{D7}}\,G_{yy}^{3/2}\,\frac{\xi}{\sqrt{\xi\,\chi-a}},
\end{align}%
where $g_{uu}^{D7}=G_{uu}+\theta'(u)$ and $\hat{G}_{1}(u)$ is
related to the $S^5$ metric \cite{Kim:2010tf}, also
\begin{subequations}\label{DCconditions1}\begin{align}
&\xi
=\tilde{E}_b^2G_3+G_{+-}G_{yy},\,\,\,\,\,\,\,\chi=G_{yy}\left(-I_y^2-\tilde{\mathcal{N}}^2\,\hat{G}_{1}(u)\,G_{yy}G_{+-}\right),\label{11}
\\&a=G_{+-}\left[\tilde{E}_b^2\left(I_+\,+\,2I_-b^2\right)^2+G_{yy}\left(I_+^2g_{++}+I_-(2I_+g_{+-}+I_-g_{--})\right)\right],\label{22}\\
&
G_{+-}=g_{++}g_{--}-g_{+-}^2,\,\,\,\,\,G_3=4b^4g_{++}-4b^2g_{+-}+g_{--},\label{33}
\end{align}\end{subequations}%
$\xi$, as a function of $u$, becomes zero at a specific point, $u_c$
where $\xi(u_c)=0$. This special point can be found by solving this
equation%
\be \tilde{E}_b^2G_3(u_c)+G_{+-}(u_c)G_{yy}(u_c)=0.\label{main1}\ee%
Reality condition of $\sqrt{\xi\,\chi-a}$ imposes two functions
$\chi$ and $a$ must vanish at $u=u_c$. Then one finds the
non-relativistic DC conductivity as follows%
\be
\sigma=2\pi\alpha'\sqrt{\frac{G_3(u_c)}{G_{yy}(u_c)}\left(\tilde{\mathcal{N}}^2G_{yy}(u_c)+\frac{I_+^2}{4\tilde{E}_b^2
b^4+g_{--}(u_c)G_{yy}(u_c)}\right)}.\label{DC1}\ee%
It is clearly seen that the non-relativistic conductivity has two
main terms \cite{Ammon:2010eq}. This is similar the relativistic
case \cite{Karch}. The term proportional to $I_+^2$ shows the
contribution from the charge carriers. The other term describes the
contribution from the charge-neutral terms which arises from
thermally produced pairs of charge carriers.

Now we find the non-relativistic conductivity in the case of the
AdS black brane where the metric functions are given by%
 \bse\begin{align}
 G_{tt}=-u^2(1-\frac{u_0^4}{u^4}),\,\,\,
 G_{yy}=u^2,\,\,\,\,
 G_{uu}=u^{-2}(1-\frac{u_0^4}{u^4})^{-1},
\end{align}\label{AdS}\ese
and the Hawking temperature in this frame is given by
\cite{Kim:2010tf}
\be\label{AdStemprature} %
 T=\frac{u_0}{\pi b}\,.
 \ee
Then \eqref{DC1} becomes \cite{Kim:2010tf}%
\be
\sigma(E,b,T)=2\pi\alpha'\sqrt{\frac{\tilde{\mathcal{N}}b^2\cos^6\theta(u_c)}{16}
\sqrt{4\tilde{E}_b^2+\pi^4T^4b^4}+\frac{4I_+^2}{4\tilde{E}_b^2+\pi^4T^4b^4}},\label{DCAdS}\ee%
Then $\sigma$ depends on the new parameter $b$ which can be
interpreted as a doping parameter \cite{KKP}. In the next section,
we more explain the importance of this parameter. It was pointed out
in \cite{ Kim:2010tf} that the expression of \eqref{DCAdS} is the
analogues of the equation (3.27) of \cite{Ammon:2010eq}. However,
they are not the same except in some limiting cases \footnote{One
may refer to \cite{Kim:2010tf} for more explanation of this point.}.
If one neglects the charge-neutral term, for weak electric field
compared
to temperature \eqref{DCAdS} becomes%
\be \sigma_1\simeq \frac{4\pi\alpha'I_+}{\pi^2 b^2 T^2}.\label{sig1}\ee%
For the opposite limit, one finds%
\be \sigma_2\simeq
2\pi\alpha'\frac{I_+}{b\tilde{E}_b}\label{sig2}.\ee It is desirable
to investigate \eqref{sig1} and \eqref{sig2} in the presence of
higher derivative corrections. However, it is not our purpose to
follow this calculation.
\section{Strange metals from AdS Light-Cone black hole}
In this section, we study the strange metal behavior in the ALCF.
Then, one should choose the gauge fields as \cite{KKP}
%
\begin{align}
  A_+=E_b y + h_+(u),\,\,\,\,\, A_-=2 b^2 E_b y+h_-(u),\,\,\, A_y=2 E_b b^2 x^- + h_y(u).\label{AAA}
\end{align}

The DBI action, then, reads%

\be S_{D7}=-\mathcal{N} \int d^5\xi \mathcal{L},
\ee%
where the Lagranigian is given by\footnote{We have assumed $G_{xx}=G_{yy}=G_{zz}$.}%
\begin{align}
  \frac{\mathcal{L}^2}{\cos^6\theta\,  G_{yy}}&= -\hat{G}\, G_{yy} G_{uu}-\tilde{E}^2_b\, \tilde{h}'^2_-
  -\tilde{E}^2_b\, g_{--} G_{uu}-\nonumber \\
  & G_{yy}\left(g_{--} \tilde{h}'^2_+-2 g_{+-}\tilde{h}'_+ \tilde{h}'_-+g_{++}\tilde{h}'^2_-
  \right)-\hat{G}\,\tilde{h}'^2_y.
\end{align}
Notice the tildes indicate that the quantities are scaled with the
factor of $2\pi\alpha'$. Also%
\be \hat{G}=-g^2_{+-}+g_{--}\,g_{++}.\ee%
The constants of motion are%
\begin{align}
  J=<J^+>=\frac{\partial \mathcal{L}}{\partial\,\tilde{h}'_+},\,\,\,\,\, <J^->=\frac{\partial \mathcal{L}}{\partial\,\tilde{h}'_-},\,\,\,\,\,
   <J^y>&=\frac{\partial \mathcal{L}}{\partial\,\tilde{h}'_y}.\label{ApApAp}
\end{align}%
Where $J,\,\, <J^->$ and $<J^y>$ are the light-cone charge density,
light-cone current along $x_-$ direction and current along $y$
direction. The boundary value of the D7-brane worldvolume field
$A_+$ acts as a source for the field theory operator $J$. The bulk
field $A_-$ is dual to the field theory operator $J_-$. In the field
theory, if one introduces $<J_+>$, one must also introduce $<J_->$
\cite{Ammon:2010eq}.

Having worked out the gauge fields in terms of these constants one
can write the on-shell Lagrangian as follows
\begin{align}\label{onshelL} %
\mathcal{L}=-&\cos^6\theta\,G_{zz}\sqrt{\hat{G}\,g_{--}\,G_{yy}\,G_{zz}
G_{uu}^{D7}}\times\nonumber\\& \left(\frac{\left( \tilde{E}^2_b
g_{--}-\hat{G}\,G_{yy}\right)^2}{\left( \tilde{E}^2_b
g_{--}-\hat{G}\,G_{yy}\right)\,\chi-\hat{G}G_{yy}G_{zz}\left(g_{--}\,<J^->+g_{+-}<J^+>\right)^2}\right)^{1/2},
\end{align}%
where $G_{uu}^{D7}=G_{uu}+\theta'(u)$ and
\begin{align}
\chi=g_{--}G_{yy}
<J^y>^2-\hat{G}\left(g_{--}G_{yy}G_{zz}\cos^6\theta+<J^+>^2\right).
\end{align}
We demand the big square root in the on-shell Lagrangian to be real
all the way from the horizon to the boundary. The explicit form of
the numerator depends on the background metric and it must be zero
somewhere between the horizon and the boundary. We assign the value
of $u$ where the numerator change the sign as $u_*$ and it can be
found by solving this equation%
 \be \left( \tilde{E}^2_b
g_{--}+\hat{G}\,G_{yy}\right)_{u=u_*}=0. \label{condition}\ee%
The denominator should also vanish at $u=u_*$. By requiring this
condition, one finds two important relations which imply that the
light-cone charge density and light-cone current are not independent as%
\begin{align}
<J^->=\left(\frac{g_{+-}}{g_{--}}\right)_{u=u_*}<J^+>,
\end{align}
, and Ohm's law, $<J^y>=\sigma E_b$, with%
\begin{align}\label{conductivity}
\sigma^2&=\sigma^{(1)}+\sigma^{(2)}\nonumber
\\ & \sim \cos^6\theta\left(\frac{\,g_{--}}{G_{yy}}\right)_{u=u_*}+\,\,J^2\left(\frac{1}
{G_{yy}^2}\right)_{u=u_*}.
\end{align}
Then the Ohmic DC conductivity is given by $\sigma$ which consists
of two terms. The important behavior of strange metals can be found
in the limit of large Light-Cone charge density $J$ \cite{ KKP}. We
consider \eqref{conductivity} in this limit and find that%
\be \rho \simeq 1/\sigma^{(2)}=\frac{G_{yy}(u_*)}{J}\ee%
Then one should find $G_{yy}(u_*)$ to investigate the temperature
dependency of the resistivity. If one considers AdS-Schwarzschild
black brane in the Light-Cone Frame, one finds that the Ohmic
resistivity depends linearly on temperature \cite{ KKP}. We show this
result in detail as follows.

The metric functions and the Hawking temperature are given in
\eqref{AdS} and \eqref{AdStemprature}. As a result, \eqref{condition} becomes%
 \be
 -u^6+u^2\,u_0^4+\frac{b^2 \tilde{E}^2_b u_0^4}{u^2}=0,
 \label{Eq0}\ee
And $u_*$ is found as%
\be u_*^2=u_0\left(\frac{u_0^2+\sqrt{4b^2
\tilde{E}^2_b+u_0^4}}{2}\right)^{1/2},\label{RAdS}\ee%
From the metric functions in \eqref{AdS}, one finds that
$G_{yy}(u_*)=u_*^2$. Then the resistivity without corrections (which is called $\rho^{(0)}$ )is given by %
\be%
\rho^{(0)}=\frac{\pi b T}{\sqrt{2}\,J}\sqrt{\pi^2 b^2 T^2+\sqrt{4b^2
\tilde{E}^2_b+\pi^4 b^4 T^4}}.
 \label{rho0}\ee
Now we consider low temperature regime of the model to study the
leading behavior of the resistivity.  In this regime
one can change $b$ at fixed $\tilde{E}_b$ and $T$. For small $T$, one finds that%
\be \rho^{(0)}=\left(\frac{\pi b \sqrt{b\tilde{E}_b}\,
}{J}\,T+\frac{\pi^3 b^3}{4\sqrt{
b\tilde{E}_b}}\,T^3+a_{2n+1}(b,\tilde{E})\,T^{2n+1}\right),\,\,\,\,\,n=3,4,5,...,
\ee%
Thus the first temperature dependence is linear. At small enough
temperatures, by increasing the parameter $b$, one finds from
\eqref{rho0} that a cross-over behavior of the resistivity from
linear to quadratic in temperature exists and one finds another
important property of the strange metals. It was argued in \cite{
KKP} that the crossover behavior is due to the fact that effectively
the gravitational background interpolates between z = 1 (AdS)
symmetry in the UV and z = 2 Lifshitz symmetry in the IR.

Now what will be happened if one considers $\mathcal{R}^4$ and $\mathcal{R}^2$
corrections to the AdS-Schwarzschild black brane? We call the
resistivity in this case as $\rho^{(\lambda)}$ and $\rho^{(\lambda_{GB})}$ and explore
effect of the finite coupling corrections on this quantity in the next section.

\subsection{The resistivity of strange metals at finite coupling}

\begin{figure}
\centerline{\includegraphics[width=75mm]{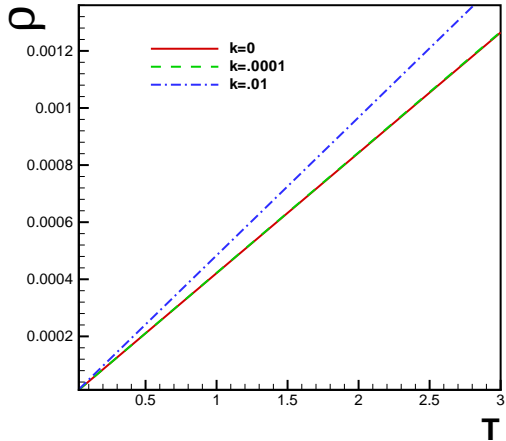} }\caption{The
resistivity of strange metals versus the temperature in the presence
of $\mathcal{R}^4$ corrections. The different values of coupling
constants are $k=0,\,k=0.0001,\, k=0.01$. Also $E=0.2$ and
$b=0.0003$.}\label{R41}
\end{figure}

Since $AdS/CFT$ correspondence refers to complete string theory, one
should consider the string corrections to the 10D supergravity
action. The first correction occurs at order $(\alpha')^3$
\cite{alpha2}. In the extremal $AdS_5\times S^5$ it is clear that
the metric does not change \cite{Banks}, conversely this is no
longer true in the non-extremal case. Corrections in inverse 't
Hooft coupling $1/\lambda$ which correspond to $\alpha^{\prime}$
corrections on the string theory side were found in \cite{alpha2}.
Functions of the $\alpha^{\prime}$-corrected metric are given by
\cite{alpha1}
\begin{eqnarray}\label{correctedmetric}
G_{tt}&=&-u^2(1-w^{-4})T(w),\nonumber\\
G_{xx}&=&u^2 X(w),\nonumber\\
G_{uu}&=&u^{-2}(1-w^{-4})^{-1} U(w),
\end{eqnarray}
where%
\begin{eqnarray}
T(w)&=&1-k\bigg(75w^{-4}+\frac{1225}{16}w^{-8}-\frac{695}{16}w^{-12}\bigg)+\dots ,\nonumber\\
X(w)&=&1-\frac{25k}{16}w^{-8}(1+w^{-4})+\dots,\nonumber\\
U(w)&=&1+k\bigg(75w^{-4}+\frac{1175}{16}w^{-8}-\frac{4585}{16}w^{-12}\bigg)+\dots,\
\end{eqnarray}
and $w=\frac{u}{u_0}$. There is an event horizon at $u=u_0$ and the
geometry is asymptotically $AdS$ at large $u$ with a radius of
curvature $R=1$. The expansion parameter $k$ can be expressed in
terms
of the inverse 't Hooft coupling as %
\be\label{k}%
 k=\frac{\zeta(3)}{8}\lambda^{-3/2}\sim 0.15\lambda^{-3/2}.
\ee %
The temperature in the light cone frame is given by%
\be %
 T_{{\cal{R}}^4}=\frac{u_0}{b \pi(1-k)}.
\ee %
By replacing \eqref{correctedmetric} to \eqref{condition}, one finds
the following equation%
\bea &&6950000 b^{32} k^3 \pi^{32} T^{32} + 6184375 b^{28} k^3
\pi^{28} T^{28} u^4 -
 125 b^{24} k^2 (71168 + 67725 k) \pi^{24} T^{24} u^8 +\nonumber\\&&
 5 b^{16} k \pi^{16} T^{16} (569344 E^2 +
    125 b^4 (1584 - 11095 k) k \pi^4 T^4) u^{12} +\nonumber\\&&
 5 b^{12} k \pi^{12} T^{12} (-632064 E^2 +
    b^4 (569344 + 1973200 k + 303125 k^2) \pi^4 T^4) u^{16} +\nonumber\\&&
 80 b^{12} k (-39664 - 12375 k + 9375 k^2) \pi^{12} T^{12} u^{20} +\nonumber\\&&
 16 b^4 \pi^4 T^4 (16 E^2 (16 + 1175 k) +
    25 b^4 (16 - 2425 k) k \pi^4 T^4) u^{24} +\nonumber\\&&
 512 b^4 (8 + 625 k) \pi^4 T^4 u^{28} - 4096 u^{32}=0, \label{Lenghty}\eea%

which is a lengthy equation. Fortunately, we are going to take the
small limit of temperature\footnote{We express the results in terms
of the temperature of hot plasma without any corrections, $
T_{\mathcal{R}^4}(1-k)=T$.},
 then we only keep terms to order
$T^8$ and rewrite \eqref{Lenghty} as%
\begin{align}
256 u^8-&32 T^4 b^4 (8+625k) \pi^4 u^4- 16 T^4 b^6 \tilde{E}_b^2
(16+1175k)+\nonumber \\ & 25 b^8 k (-16+2425k) \pi^8 T^8=0.
\label{Eq4}\end{align}%
This equation should be solved to find $u_*$. It is clearly seen
that \eqref{Eq4} is the same as the equation of \eqref{Eq0}. Then
one expects the same solution while the $\mathcal{R}^4$ corrections
have been included\footnote{ It should be noticed that we are
interested
in the case of small temperature regime.}. The final expansion of the resistivity at finite coupling as follows%
\be \rho^{(\lambda)}=\hat{a}_{1}(b, E, k)\, T^{}+\hat{a}_{3}(b, E,
k)\, T^{3}+\hat{a}_{2n+1}(b, E, k)\, T^{2n+1},\,\,\,n=0,1,2,3,...
\ee%
where%
\begin{align}
\hat{a}_1&=\frac{1}{2}\left( 16+1175k\right)^{1/4}\,\left(\frac{\pi b \sqrt{b\tilde{E}_b}}{J\,}\right),\nonumber \\
\hat{a}_3&=\frac{b^3 \pi^3 (8+575k)}{16 J \left(b^2
\tilde{E}_b^2(16+1175k)\right)^{1/4}}.
\end{align}%
It can be seen that the result obtained in \cite{KKP} can be
reproduced even at higher-order corrections. This is an interesting
result that $\mathcal{R}^4$ corrections keep the linear temperature
dependent resistivity and at the leading
order enhance the coefficient of it as%
\be \hat{a}_1=\frac{\pi b \sqrt{b\tilde{E}_b}}{2\,J}\left( 16+176.25
\lambda^{-3/2}\right)^{1/4}.
\ee%
To support our conclusion, we use the numerical methods and plot the
resistivity in the presence of $\mathcal{R}^4$ corrections in Fig.
\ref{R41}. We assume $E=.2$ and $b=.0003$. For small values of $k$,
for example $k=.0001$, the effect of corrections are negligible. It
is clearly seen that by increasing the parameter $k$, the slope of
line is changing.

Our numerical calculations also supports that by increasing the bulk
parameter $b$, the temperature dependence of the resistivity crosses
from linear to quadratic. This is another interesting property of
strange metals which can be found from AdS Light-Cone black hole and
is also valid in the presence of $\mathcal{R}^4$ corrections.

Next, we study $\mathcal{R}^2$ corrections to the resistivity and
call it $\rho^{(\lambda_{GB})}$. In five dimensions, we consider the
theory of gravity with quadratic powers of curvature as
Gauss-Bonnet(GB) theory. The exact solutions and thermodynamic
properties of the black brane in GB gravity were
discussed in \cite{Cai:2001dz,Nojiri:2001aj,Nojiri:2002qn}. The metric functions are given by%
\begin{equation}
G_{tt}=-N \,u^2\, h(u),\,\,\,\,\,\, G_{uu}=\frac{1}{u^2
h(u)},\,\,\,\,\, G_{xx}=G_{yy}=G_{zz}=u^2\label{GBmetric},
\end{equation}
where
\begin{equation}
h(u)= \frac{1}{2\lambda_{GB}}\left[ 1-\sqrt{1-4 \lambda_{GB}\left(
1-\frac{u_0^4}{u^4} \right)}\right].
\end{equation}
In (\ref{GBmetric}), $N= \frac{1}{2}\left(
 1+\sqrt{1-4 \lambda_{GB}} \right)$ is an arbitrary constant which specifies
the speed of light of the boundary gauge theory and we choose it to
be unity. The temperature is given by
\begin{equation}
 T_{GB}=\frac{\sqrt{N}\,\, u_0}{b \pi \,}.
\end{equation}
Then equation \eqref{condition} reads%
\be \frac{u^2}{2 \lambda_{GB}}\left(2 b^2 \lambda_{GB}
\tilde{E}_b^2+ N(u^4+b^2 \tilde{E}_b^2)\left(-1+\sqrt{1+4
\lambda_{GB}\left(-1+\frac{u_0^4}{u^4}\right)}\right)\right)=0.
\ee%
One should solve the above equation to find curvature-squared
corrections to the Ohmic resistivity as follows\footnote{We also
express here the results in terms of the temperature of hot plasma
without any corrections, $
\frac{T_{GB}}{\sqrt{N}}=T$.}%

\bea 6\,J\,\sqrt{ \rho^{\lambda_{GB}}}&=&
\frac{b^4T^4}{N^2}-\frac{2(1-2 N)b^2 \tilde{E}_b^2}{N} +
\nonumber\\&& 2\,2^{1/3}\left(b^4
\tilde{E}_b^4(1-N+N^2-3\lambda_{GB})+ \frac{2(1+N)}{N}b^6
\tilde{E}_b^2\,T^4+\frac{\,b^8 T^8}{N^2}\right)\times \nonumber\\&&
g(\tilde{E}_b,\lambda_{GB},b)^{-1/3} +\nonumber\\&&
\frac{2\, 2^{2/3}}{N^2}\left(I(\lambda_{GB},E_b,b)+3\sqrt{3} N^3b^3E_b^3\sqrt{h(\lambda_{GB},E_b,b)}\right)^{1/3}, \eea%
where%
\bea g(\tilde{E}_b,\lambda_{GB},b)&=&N^3 b^6 E_b^6(9
\lambda_{GB}-2)-2N^6(b^2
E_B^2+\frac{b^4T^4}{N^2})^3+\nonumber\\&&3N^4b^4 E_b^4\left(b^2
E_b^2(1-6
\lambda_{GB})+(3\lambda_{GB}-2)\frac{b^4T^4}{N^2}\right)+\nonumber\\&&
3N^5\left(b^6
E_b^6-b^4E_b^4-\frac{2b^{10}E_b^2T^8}{N^4}\right)+\nonumber\\&&
3\sqrt{3} N^3b^3E_b^3\sqrt{h(\lambda_{GB},E_b,b)},\nonumber\\
\nonumber\\
h(\lambda_{GB},E_b,b)&=&b^6E_b^6
\lambda_{GB}^2(4\lambda_{GB}-1)-\nonumber\\&& 2N
b^4E_b^4\lambda_{GB}(b^2E_b^2(-1+4\lambda_{GB})+\frac{\lambda_{GB}b^4T^4}{N^2})+\nonumber\\&&
N^4 (b^2 E_b^2 + \frac{B^4T^4}{N^2})^2 \left(b^2 E_b^2 (-1 +
4\lambda_{GB} ) + 4 \frac{\lambda_{GB}
b^4T^4}{N^2}\right)+\nonumber\\&& 2 N^3 b^2 E_b^2 \left(b^4 E_b^4 (1
- 4 \lambda_{GB}) + \frac{b^6 T^4 E_b^2}{N^2} (-1 + 4\lambda_{GB}) +
5 \frac{\lambda_{GB}b^8T^8}{N^4}\right)+\nonumber\\&& N^2 (b^6 E_b^6
(-1 + 2 \lambda_{GB} + 8 \lambda_{GB}^2)+ 4\frac{ b^8 T^4
E_b^4}{N^2} \lambda_{GB} (2 -
5 \lambda_{GB})- \frac{b^{10}T^8 E_b^2}{N^4} \lambda_{GB}^2 ),\nonumber \eea%
and%
\bea
I(\lambda_{GB},E_b,b)&=&N^3b^6E_b^6(9\lambda_{GB}-2)-2N^6(b^2E_b^2+\frac{b^4T^4}{N^2})^3+\nonumber\\&&
3N^4b^4E_b^4\left(b^2E_b^2(1-6\lambda_{GB})+(-2+3\frac{\lambda_{GB})b^4T^4}{N^2}\right)+\nonumber\\&&
3N^5\left(b^6E_b^6-b^4E_b^8T^4-\frac{2b^{10}E_b^2T^8}{N^4}\right).
\eea%

At the limit of large density and low temperature, one finds the
following expansion%
\be \rho^{(\lambda_{GB})}=A_0(b, \lambda_{GB},E_b)+A_{4}(b, \lambda_{GB},E_b)\,T^4+ A_{8}(b, \lambda_{GB},E_b)\, T^8+...,\ee%
It is clearly seen that the Ohmic resistivity can never be linear in
the case of $\mathcal{R}^2$ corrections. This is an interesting
result and one may define a new class of strange metals where the
resistivity depends on $T^4$.\footnote{Certainly, this class must be
study more in details and other properties should be investigated.}

Now, We increase the doping parameter b to study behavior of the resistivity at finite coupling.
This limit was studied in details in \cite{KKP}. In this paper a dimensionless variable was defined as%
\be t=\frac{\pi R T b}{\sqrt{2b \tilde{E}_b}}. \ee %
Different limits of resistivity was studied as $t\gg 1$ and $t \ll
1$ which correspond to $b T \gg \sqrt{2 b\tilde{E}_b}$ and $b T \ll
\sqrt{2 b\tilde{E}_b}$, respectively. The parameters $b$ is an extra
parameter which exists in ALCF geometry. It can be identified as
doping parameter. Then by increasing $b$, one may neglect the
electric field and by increasing it the electric field becomes
strong \cite{Kim:2011zd}. Now we follow these limits.

By increasing $b$, we neglect the electric field which means
$\tilde{E}_b\rightarrow 0$ . As a result \eqref{condition} is
simplified to $\hat{G}\,G_{yy}=0$. The specific solution of this
equation occurs at the horizon $u_*=u_0$. For example in
\eqref{Eq0}, by applying this limit one finds that $u_*\rightarrow
u_0$. This limit is also valid in the presence of higher derivative
corrections, i.e $\mathcal{R}^4$ and $\mathcal{R}^2$. This is
another important property of strange metals that by increasing the
doping parameter, one
 finds a cross-over from linear to quadratic in temperature. We find that higher derivative corrections to the resistivity keep
 this property and  show the cross over to quadratic in temperature. This is an interesting result which shows
that one of the important properties of strange metals can be
achieved by AdS Light-Cone approach even in the presence of higher
derivative corrections.
\subsection{Hall conductivity of strange metals at finite coupling}
The inverse Hall angle is defined as follows%
\be \frac{\sigma^{yy}}{\sigma^{yz}},\ee %
where $\sigma^{yy}$ and $\sigma^{yz}$ are Ohmic and Hall conductivity, respectively. The temperature dependence
$( \sim T^2)$ of the inverse Hall angle is the typical property of the strange metal. To study the Hall conductivity of strange metals, one should turn on a magnetic field.
We follow \cite{KKP,Kim:2011zd} and introduce the gauge fields as%
\be
\tilde{A}=(\tilde{E}_b\,y+h_+(u))dx^++h_-(u)dx^-+h_y(u)dy+(\tilde{B}_b\,y)dz.
\ee%
It implies that there is an electric field, $E_b$, along
the $y$ direction and a magnetic field, $B_b$, along the $z$
direction. The on-shell DBI action is %
\begin{align}%
S_{D7}=-\tilde{\mathcal{N}}^2\, \int du\,
\sqrt{G_{+-}\,g_{--}\,g_{uu}^{D7}}\,G_{yy}^{}\,\frac{\xi_1}{\sqrt{\xi_1\,\chi_1-G_{yy}^2\,G_{+-}\,a_1^2-a_2^2}},
\end{align}%
where $g_{uu}^{D7}=G_{uu}+\theta'(u)$ and $\hat{G}_{i}(u)$ is
related to the $S^5$ metric \cite{Kim:2010tf}, also
\begin{subequations}\label{DCconditions1}\begin{align}
&\xi_1
=\tilde{E}_b^2\,G_{yy}\,g_{--}-G_{+-}\left(G_{yy}^2+\tilde{B}_b^2\right),
\label{00}\\&\chi_1=-G_{+-}\left(g_{--}G_{yy}^2\cos(\theta)^3+I_+^2\right)+g_{--}G_{yy}(I_{z}^2+I_y^2),\label{11}
\\&a_1=g_{--}I_-+g_{+-}I_+,\,\,\,\,\,a_2=g_{--}G_{yy}\tilde{E}_b I_z+G_{+-}\,\tilde{B}_b\,I_+,\label{22}\\
& G_{+-}=g_{++}g_{--}-g_{+-}^2,\label{33}
\end{align}\end{subequations}%
Reality condition imposes that $\xi_1(u_c)=0$. Then one should solve the following equation%
\be\tilde{E}_b^2\,G_{yy}\,g_{--}-G_{+-}\left(G_{yy}^2+\tilde{B}_b^2\right)=0, \label{MainHall}\ee  The Ohmic conductivity in the presence of magnetic field is given by%
\be \sigma^{yy}=\frac{\sqrt{G_{yy}g_{--}\cos(\theta)^3\left(G_{yy}^2+\tilde{B}_b^2\right)+I_+^2}}
{G_{yy}^2+\tilde{B}_b^2}, \ee %
and the Hall conductivity is%
\be \sigma^{yz}=\frac{\tilde{B}_b I_+}{G_{yy}^2+\tilde{B}_b^2}. \ee %
These conductivities must be evaluated at $u=u_c$ which can be found from \eqref{MainHall}. At large density and weak magnetic field, one finds that%
\be \sigma^{yy}\sim\frac{I_+}{G_{yy}(u_c)},\,\,\,\,\,\sigma^{yz}=
\frac{\tilde{B}_b I_+}{G_{yy}(u_c)^2},\ee %
As a result, the inverse Hall conductivity is given by the following expression%
\be \frac{\sigma^{yy}}{\sigma^{yz}}\sim \frac{G_{yy}(u_c)}{\tilde{B}_b}, \ee %
 Then simply one should find $G_{yy}(u_c)$. To investigate the strange metal behavior, one should consider special limits of magnetic filed, electric field and temperature. Fortunately, ALCF benefits from extra parameter $b$. To more clarify the role of this parameter, we solve \eqref{MainHall} in the case of AdS spacetime withought any corrections. This case was studied in \cite{KKP,Kim:2011zd}.
From the metric functions in \eqref{AdS}, one finds%
\be -u^8+ \tilde{E_b}^2 b^6T^4+u^4 b^4 T^4+\tilde{B_b}^2(b^4T^4-u^4)=0.\label{Halleq}\ee %
The analytic solution of the above equation would be easily found.
However, we are going to understand different limits of parameters
in this equation and apply it in the case of complicated equation
which appears in higher derivative corrections.
We consider different limits as follows%
\begin{itemize}
\item{ $\tilde{B}_b=0$, by increasing the doping parameter ($\tilde{E}_b\rightarrow 0$), one may neglect the electric field. In this case $u_c=b T$ then
$G_{yy}(u_c)=u_c^2 \sim T^2$ }
\item{$\tilde{B}_b=0$, by decreasing the doping parameter ($\tilde{E}_b\rightarrow \infty$), one may keep only the term which depends on the electric field. In this case $u_c \sim T$ then
$G_{yy}(u_c)=u_c^2 \sim T$}
\item{$\tilde{B}_b\neq 0$, one may assume $\tilde{B}_b > b\tilde{E}_b$ and decrease the doping parameter. In this case, one should
 only keep the terms which are depended on the magnetic field and \eqref{Halleq} becomes $\tilde{B}_b^2(b^4T^4-u^4)=0$. Then
 $u_c=b T$ and $G_{yy}(u_c) \sim T^2$.      }
\end{itemize}
Now, one finds that in the presence of higher derivative corrections, the behavior of the inverse Hall effect will not be changed.
Because in the limit of nonezero magnetic field and  $\tilde{B} > \tilde{E}$ by decreasing the doping parameter, \eqref{MainHall} becomes%
\be -G_{+-} \tilde{B}_b^2=0.\ee %
Where simply the solution is $u_c=u_0$.
\section{Conclusion}
It is well known that the light cone quantization of a relativistic
theory looks like a non-relativistic theory. As a simple holographic
system, one may use the AdS-Schwarzshild black hole in light-con
coordinates to investigate strange metal properties in
\eqref{strange}. The universal experimental results of strange
metals were studied in \cite{KKP}. The non-relativistic DC
conductivity is particularly
important which is characterized by%
\be \sigma \sim \frac{1}{T} \label{1}\ee%
As it is clear the resistance $\rho = \frac{1}{\sigma}$ increases
linearly with the temperature, this is the reason why 'strange' is
defined \cite{Sachdev:2010uj}. One should notice that \eqref{1} is
valid at low temperatures compared to the mass and density. In this
paper, we have studied the effects of finite but large couplings by
adding higher-derivative corrections in the gravity background.
Especially, $\mathcal{R}^4$ terms and Gauss-Bonnet gravity has been
considered. It is found that the Ohmic resistivity and inverse Hall
angle are linear and quadratic temperature dependent in the case of
$\mathcal{R}^4$ corrections, respectively. While Ohmic conductivity
can never be linear temperature dependent and $\rho=
A_0+A_4T^4+A_8T^8$. \footnote{$A_0$,$A_4$ and $A_8$ depend on the
electric field, Gauss-Bonnet coupling and parameter $b$.} This is an
interesting result and may be define a new class of strange metals.

\section*{Acknowledgment}
We would like to thank M. Ali-Akbari, M. Alishahiha and M.
Sheikh-Jabbari for very useful discussions and especially thank Bom
Soo Kim for reading the manuscript and useful comments. This
research was supported by Shahrood University of Technology.


\begin{thebibliography}{99}
\bibitem{review1}
S. A. Hartnoll, "Lectures on holographic methods for condensed
matter physics," Class.Quant.Grav. 26 (2009) 224002,
[arXiv:0903.3246].\\
J. McGreevy, "Holographic duality with a view toward many-body
physics," Adv.High Energy Phys. 2010 (2010) 723105,
[arXiv:0909.0518].\\
S. A. Hartnoll, "Horizons, holography and condensed matter,"
arXiv:1106.4324. \\
N.~Iqbal, H.~Liu and M.~Mezei,
  ``Lectures on holographic non-Fermi liquids and quantum phase transitions,''
  arXiv:1110.3814 [hep-th].

\bibitem{strange1}
G. R. Stewart, "Non-Fermi-liquid behavior in d- and f-electron
metals," Rev. Mod. Phys. 73, 797 (2001) [Addendum-ibid. 78, 743
(2006)].

\bibitem{strange2}
R. A. Cooper, Y. Wang, B. Vignolle, O. J. Lipscombe, S. M. Hayden,
Y. Tanabe, T. Adachi, Y. Koike, M. Nohara, H. Takagi, C. Proust, and
N. E. Hussey, "Anomalous criticality in the electrical resistivity
of La2-xSrxCuO4," Science, 323, 603 (2009).

\bibitem{Sachdev:2010uj}
  S.~Sachdev,
  ``Strange metals and the AdS/CFT correspondence,''
  J.\ Stat.\ Mech.\  {\bf 1011} (2010) P11022
  [arXiv:1010.0682 [cond-mat.str-el]].



\bibitem{Son:2008ye}
  D.~T.~Son,
  ``Toward an AdS/cold atoms correspondence: A Geometric realization of the
  Schrodinger symmetry,''
  Phys.\ Rev.\  D {\bf 78} (2008) 046003
  [arXiv:0804.3972 [hep-th]].
\bibitem{Balasubramanian:2008dm}
  K.~Balasubramanian and J.~McGreevy,
  ``Gravity duals for non-relativistic CFTs,''
  Phys.\ Rev.\ Lett.\  {\bf 101} (2008) 061601
  [arXiv:0804.4053 [hep-th]].
  \bibitem{Maldacena}
  J.~Maldacena, D.~Martelli and Y.~Tachikawa,
  ``Comments on string theory backgrounds with non-relativistic conformal
  symmetry,''
  JHEP {\bf 0810} (2008) 072
  [arXiv:0807.1100 [hep-th]].
\bibitem{Alishahiha:2003ru}
  M.~Alishahiha and O.~J.~Ganor,
  ``Twisted backgrounds, PP waves and nonlocal field theories,''
  JHEP {\bf 0303} (2003) 006
  [arXiv:hep-th/0301080].
\bibitem{Goldberger:2008vg}
  W.~D.~Goldberger,
  ``AdS/CFT duality for non-relativistic field theory,''
  JHEP {\bf 0903} (2009) 069
  [arXiv:0806.2867 [hep-th]].\\
  J.~L.~F.~Barbon and C.~A.~Fuertes,
  ``On the spectrum of nonrelativistic AdS/CFT,''
  JHEP {\bf 0809} (2008) 030
  [arXiv:0806.3244 [hep-th]].
\bibitem{Kim:2010tf}
  B.~S.~Kim and D.~Yamada,
  ``Properties of Schroedinger Black Holes from AdS Space,''
  JHEP {\bf 1107} (2011) 120
  [arXiv:1008.3286 [hep-th]].

\bibitem{Faulkner:2010zz}
  T.~Faulkner, N.~Iqbal, H.~Liu, J.~McGreevy and D.~Vegh,
  ``Strange metal transport realized by gauge/gravity duality,''
  Science {\bf 329} (2010) 1043.

\bibitem{Charmousis:2010zz}
  C.~Charmousis, B.~Gouteraux, B.~S.~Kim, E.~Kiritsis and R.~Meyer,
  ``Effective Holographic Theories for low-temperature condensed matter systems,''
  JHEP {\bf 1011} (2010) 151
  [arXiv:1005.4690 [hep-th]].
\bibitem{Myers:2010pk}
  R.~C.~Myers, S.~Sachdev and A.~Singh,
  ``Holographic Quantum Critical Transport without Self-Duality,''
  Phys.\ Rev.\ D {\bf 83} (2011) 066017
  [arXiv:1010.0443 [hep-th]].
\bibitem{Pal:2010sx}
  S.~S.~Pal,
  ``Model building in AdS/CMT: DC Conductivity and Hall angle,''
  Phys.\ Rev.\ D {\bf 84} (2011) 126009
  [arXiv:1011.3117 [hep-th]].

\bibitem{Lee:2010ii}
  B.~-H.~Lee, D.~-W.~Pang and C.~Park,
  ``Strange Metallic Behavior in Anisotropic Background,''
  JHEP {\bf 1007} (2010) 057
  [arXiv:1006.1719 [hep-th]].
\bibitem{Meyer:2011xn}
  R.~Meyer, B.~Gouteraux and B.~S.~Kim,
  ``Strange Metallic Behaviour and the Thermodynamics of Charged Dilatonic Black Holes,''
  Fortsch.\ Phys.\  {\bf 59} (2011) 741
  [arXiv:1102.4433 [hep-th]].

\bibitem{Lee:2010uy}
  B.~-H.~Lee and D.~-W.~Pang,
  ``Notes on Properties of Holographic Strange Metals,''
  Phys.\ Rev.\ D {\bf 82} (2010) 104011
  [arXiv:1006.4915 [hep-th]].
\bibitem{pol}
S.~A.~Hartnoll,~J.~Polchinski,~E.~Silverstein and D.~Tong, ``Towards
strange Metallic holography,''


\bibitem{KKP}
  B.~S.~Kim, E.~Kiritsis and C.~Panagopoulos,
  ``Holographic quantum criticality and strange metal transport,''
  New J.\ Phys.\  {\bf 14} (2012) 043045
  [arXiv:1012.3464 [cond-mat.str-el]].

\bibitem{Kim:2011zd}
  K.~-Y.~Kim, D.~-W.~Pang,
  ``Holographic DC conductivities from the open string metric,''
  JHEP {\bf 1109 } (2011)  051.
  [arXiv:1108.3791 [hep-th]].

\bibitem{AliAkbari:2010av}
  M.~Ali-Akbari and K.~B.~Fadafan,
  ``Conductivity at finite 't Hooft coupling from AdS/CFT,''
  arXiv:1008.2430 [hep-th].


\bibitem{Ammon:2010eq}
  M.~Ammon, C.~Hoyos, A.~O'Bannon and J.~M.~S.~Wu,
  ``Holographic Flavor Transport in Schrodinger Spacetime,''
  JHEP {\bf 1006} (2010) 012
  [arXiv:1003.5913 [hep-th]].


\bibitem{Karch}
  A.~Karch and A.~O'Bannon,
  `Metallic AdS/CFT,''
  JHEP {\bf 0709} (2007) 024
  [arXiv:0705.3870 [hep-th]].


\bibitem{alpha2}
 J. Pawelczyk and S. Theisen,
 {\it AdS$_5\times S^5$ black hole metric at {\cal O}($\alpha^{\prime 3}$)},
  JHEP {\bf 9809} (1998) 010, [hep-th/9808126];

\bibitem{Banks}
  T.~Banks and M.~B.~Green,
  ``Non-perturbative effects in AdS(5) x S**5 string theory and d = 4 SUSY
  Yang-Mills,''
  JHEP {\bf 9805}, 002 (1998)
  [arXiv:hep-th/9804170];

\bibitem{alpha1}
 S.S. Gubser, I.R. Klebanov and A.A. Tseytlin,
{ \it Coupling constant dependence in the thermodynamics of $N=4$
supersymmetric Yang-Mills theory}
 Nucl. Phys. {\bf B534} (1998) 202, [hep-th/9805156];

\bibitem{Cai:2001dz}
  R.~G.~Cai,
  ``Gauss-Bonnet black holes in AdS spaces,''
  Phys.\ Rev.\  D {\bf 65} (2002) 084014
  [arXiv:hep-th/0109133].
\bibitem{Nojiri:2001aj}
  S.~Nojiri and S.~D.~Odintsov,
  ``Anti-de Sitter black hole thermodynamics in higher derivative gravity  and
  new confining-deconfining phases in dual CFT,''
  Phys.\ Lett.\  B {\bf 521} (2001) 87
  [Erratum-ibid.\  B {\bf 542} (2002) 301]
  [arXiv:hep-th/0109122].
\bibitem{Nojiri:2002qn}
  S.~Nojiri and S.~D.~Odintsov,
  "(Anti-) de Sitter black holes in higher derivative gravity and dual
  conformal field theories,"
  Phys.\ Rev.\  D {\bf 66} (2002) 044012
  [arXiv:hep-th/0204112].
\end{thebibliography}
\end{document}